\documentclass[aip,rsi,reprint]{revtex4-1}
\usepackage{amssymb}
\usepackage{amsmath}
\usepackage{textcomp}
\usepackage{graphicx} 
\usepackage{epstopdf}
\usepackage{dcolumn}
\usepackage{bm}
\usepackage{soul}
\usepackage{color}
\usepackage{hyperref}
\hypersetup{colorlinks=true,linkcolor=red,citecolor=red,urlcolor=red}
\usepackage{upgreek}
\usepackage{soul}

\def\um{\upmu\mbox{m}}

\def\Wcm2{\mbox{W cm}^{-2}}
\def\Wcmum2{\mbox{Wcm}^{-2}\upmu\mbox{m}^{2}}

\def\cm3{\mbox{cm}^{-3}}

\begin{document}

\preprint{AIP/123-QED}

\title{Calibration of Time Of Flight Detectors Using Laser-driven Neutron Source}

\author{S. R. Mirfayzi}
\affiliation{Centre for Plasma Physics, School of Mathematics and Physics, Queen's University Belfast, BT7 1NN, UK}

\author{S.~Kar}\email{s.kar@qub.ac.uk}
\affiliation{Centre for Plasma Physics, School of Mathematics and Physics, Queen's University Belfast, BT7 1NN, UK}

\author{H.~Ahmed}
\affiliation{Centre for Plasma Physics, School of Mathematics and Physics, Queen's University Belfast, BT7 1NN, UK}

\author{A.G.~Krygier}
\affiliation{Department of Physics, The Ohio State University, Columbus, Ohio 43210, USA}

\author{A.~Green}
\affiliation{Centre for Plasma Physics, School of Mathematics and Physics, Queen's University Belfast, BT7 1NN, UK}

\author{A.~Alejo}
\affiliation{Centre for Plasma Physics, School of Mathematics and Physics, Queen's University Belfast, BT7 1NN, UK}

\author{R.~Clarke}
\affiliation{Central Laser Facility, Rutherford Appleton Laboratory, Didcot, Oxfordshire, OX11 0QX, UK}

\author{R.R.~Freeman}
\affiliation{Department of Physics, The Ohio State University, Columbus, Ohio 43210, USA}

\author{J.~Fuchs}
\affiliation{LULI, Ecole Polytechnique, CNRS, Route de Saclay,
91128 Palaiseau Cedex,France}

\author{D.~Jung}
\affiliation{Centre for Plasma Physics, School of Mathematics and Physics, Queen's University Belfast, BT7 1NN, UK}

\author{A. Kleinschmidt}
\affiliation{Institut f\"ur Kernphysik, Technische Universit\"at Darmstadt, Schlo{\ss}gartenstrasse 9, D-64289 Darmstadt, Germany}

\author{J.T.~Morrison}
\affiliation{Propulsion Systems Directorate, Air Force Research Lab, Wright Patterson Air Force Base, Ohio 45433, USA}

\author{Z.~Najmudin}
\affiliation{Blackett Laboratory, Department of Physics, Imperial College, London SW7 2AZ, UK}

\author{H.~Nakamura}
\affiliation{Blackett Laboratory, Department of Physics, Imperial College, London SW7 2AZ, UK}

\author{P.~Norreys}
\affiliation{Central Laser Facility, Rutherford Appleton Laboratory, Didcot, Oxfordshire, OX11 0QX, UK}
\affiliation{Department of Physics, University of Oxford, Oxford, OX1 3PU, UK}

\author{M.~Oliver}
\affiliation{Department of Physics, University of Oxford, Oxford, OX1 3PU, UK}

\author{M.~Roth}
\affiliation{Institut f\"ur Kernphysik, Technische Universit\"at Darmstadt, Schlo{\ss}gartenstrasse 9, D-64289 Darmstadt, Germany}

\author{L.~Vassura}
\affiliation{LULI, Ecole Polytechnique, CNRS, Route de Saclay,
91128 Palaiseau Cedex,France}
%\affiliation{Dipartmento SBAI, Universit\'a di Roma �La Sapienza, 00161 Rome, Italy}

\author{M.~Zepf}
\affiliation{Centre for Plasma Physics, School of Mathematics and Physics, Queen's University Belfast, BT7 1NN, UK}
\affiliation{Helmholtz Institut Jena, D-07743 Jena, Germany}

\author{M.~Borghesi}
\affiliation{Centre for Plasma Physics, School of Mathematics and Physics, Queen's University Belfast, BT7 1NN, UK}
\affiliation{Institute of Physics of the ASCR, ELI-Beamlines project, Na Slovance 2, 18221 Prague, Czech Republic}

\date{\today}% It is always \today, today,

\begin{abstract}
Calibration of three scintillators (EJ232Q, BC422Q and EJ410) in a time-of-flight (TOF) arrangement using a laser drive-neutron source is presented. The three plastic scintillator detectors were calibrated with gamma insensitive bubble detector spectrometers, which were absolutely calibrated over a wide range of neutron energies ranging from sub MeV to 20 MeV. 
A typical set of data obtained simultaneously by the detectors are shown, measuring the neutron spectrum emitted from a petawatt laser irradiated thin foil.

%Valid PACS numbers may be entered using the \verb+\pacs{#1}+ command.
\end{abstract}

\maketitle

\section{\label{sec:level1}Introduction}
\begin{figure*}
\includegraphics[width=.25\textwidth, angle =-90]{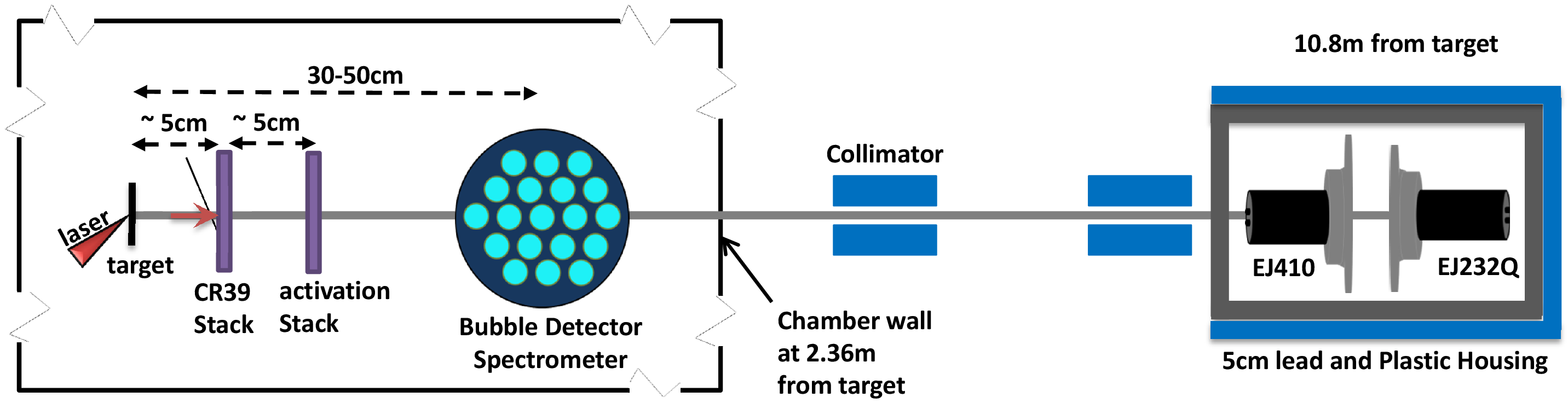}
\caption{\label{fig:epsart1} Schematic of the experimental setup. The neutron beam generated from laser irradiated target or catcher propagates through various diagnostics before reaching the scintillator detectors (Figure not drawn to scale). The scintillator detectors were placed face-on to each other inside a lead and plastic housing. The collimators, made of 5 cm thick plastic, were used before the detectors in order to reduce stray neutrons hitting the detector.}
\end{figure*}

In recent years, there has been a growing interest in laser-driven sources for their high brightness, directionality and compactness, which have shed a new light on the production of ultra-short (sub-ns) neutron pulses that can be used in many potential applications in science\cite{roth2013bright}, industry\cite{perkins2000investigation}, security\cite{Loveman1995765} and healthcare\cite{borghesi2006fast,macchi2013ion}. Since laser-driven neutron sources are primarily based on beam-fusion or spallation reactions initiated by fast ions\cite{mckenna2005broad}, the measurement of absolute neutron spectrum is crucial not only for the development and optimisation of the neutron sources, but also for the study of the parent ions involved in the neutron generation process.

Among commonly used neutron diagnostics, those based on time-of-flight (TOF) arrangement are probably the most widely used for characterising the neutron spectra. Depending on the distance between the source and the detector, a TOF spectrometer can in principle resolve neutron energies ranging from sub-MeV to hundreds of MeV. However, the most unique feature of TOF scintillator detector lies in the capability to provide an absolute spectral measurements of fast neutrons, which are otherwise first moderated to lower energies in other types of neutron spectrometers resulting in a poor spectral information\cite{harvey1979scintillation}. Most common TOF detectors are made of scintillators coupled to photo-multiplier tubes (PMT), where the specifications of the scintillators and PMTs depend on the type and range of measurements sought. Plastic scintillators are the most common type of detectors employed for neutron detection due to their high dynamic range and ease of handling.

In order to obtain the complete spectral information from a TOF diagnostic, one would require an absolute calibration of the detector over a range of neutron energies. Traditionally, the scintillators are calibrated with continuous sources, such as $\textsuperscript{252}$Cf emitting neutrons via spontaneous fission reactions~\cite{green1973californium,mannhart1986evaluation}, by using pulse-height discrimination technique in multi-channel analyser\cite{verbinski1968calibration}. However, the radioactive isotopes do not provide an accurate measurement of detector response due to their broad energy spectrum (for example, 0.1 - 10 MeV for $\textsuperscript{252}$Cf).

Since the scintillators fluoresce by the knock-on protons produced by neutron elastic scattering, their response to different energy neutrons exhibit a highly non-linear behaviour. 
In this paper we report on an alternative way of calibrating the scintillator detectors by using an ultra-short burst of neutrons, which enables an energy resolved calibration by deploying the detector in a TOF arrangement. High power laser driven pulsed neutron sources emit high flux of neutrons within a short burst duration (sub-ns). Therefore using a detector at a suitably large distance (several meters) from the source temporally disperses the arrival of different energy neutrons on the detector. This provides an ideal situation for energy resolved calibration of the detector using a broadband source. This is unlike the pulse-height discrimination technique, where the neutron energy is identified from the height of the pulses generated by neutron induced scintillations, while assuming that each pulse is produced by a single hit of neutron. The laser driven source used in our case delivered high flux of neutrons of the order of $10^9$ n/sr in a single shot, which was able to produce a strong signal in the TOF detector for cross-calibration with the absolutely calibrated bubble detector spectrometers (BDS) fielded in the same shot. 

The calibration technique presented here combines a series of experimental measurements and Monte-Carlo simulations, which was employed to calibrate three detectors made of different plastic scintillators. As a typical detector consists of many parts, such as scintillator, light guide and photo-multiplier tube (PMT), an absolute calibration requires specific knowledge of the detector design and characteristics of its different parts, such as the detector housing, light guide, PMT etc. However, the non-linear response of the detector to different energy neutrons comes from the plastic scintillator itself, which originates from the difference in recoil proton spectra produced by different energy neutrons. The relative response of the scintillator over a range of neutron energy was obtained by monte-carlo simulations, while benchmarking the results with the literature. The neutron spectra from the TOF data was obtained by using the energy dependent scintillator response, while taking into account the transmission of different energy neutrons through various objects along the line of sight to the detector. Finally, the detectors were calibrated by comparing the TOF neutron spectra with that obtained by the absolutely calibrated BDS spectrometers employed in the same shot.  

The subject matter in this paper is organised in four sections. Section \ref{sec1:levelii} illustrates a brief description of the experimental setup. Section \ref{sec:working} describes the basic principles of neutron detection by scintillation processes. Section \ref{sec2:leveliv} outlines the numerical work carried out by employing Monte-Carlo simulations in order to obtain the relative response of our scintillator to different energy of fast neutrons.
Section \ref{sec2:levelv} reports the final calibration obtained for the three scintillator detectors and typical neutron spectra obtained by different detectors.   

\section{\label{sec1:levelii}Experimental Setup}%\protect
Calibration of the scintillator detectors was carried out using a laser driven neutron beam generated at VULCAN Petawatt target area of Rutherford Appleton Laboratory, STFC, UK\cite{danson2004vulcan}. The 750fs FWHM laser pulse of Vulcan, with energy of 600J, was focused on the target by a f/3 off-axis parabola delivering intensity in the range of $1-5 \times 10\textsuperscript{20}Wcm^{-2}$ . Energetic ions were produced by irradiating the laser on deuterated plastic foils of various thickness, which were employed to generate neutrons through beam fusion reaction by interacting with either the bulk target ions or secondary $C_2D_4$ catcher target placed at a small distance ($\sim$cm) from the laser irradiated target.   

A suite of diagnostics, such as scintillator TOF, BDS, activation samples\cite{wolle1999neutron, cooper2001nif}  and CR39 stacks\cite{seguin2003spectrometry,sinenian2011response,bang2012calibration} were employed around the target in order to characterise the neutron beam parameters. A schematic of a typical experimental setup used for cross-calibrating the scintillator detectors with the bubble detectors is shown in Fig.~\ref{fig:epsart1}. The bubble detectors, manufactured and calibrated by BTI Bubble technology industries\cite{BTI}, were placed inside the interaction chamber at various distances (30-50 cm) from the neutron source in order to ensure a sufficient number (at least $\sim$ 50 bubbles per energy bin) of bubble formation. The scintillator detectors were placed outside the interaction chamber at a significantly larger distance from the neutron source to achieve high energy resolution by the TOF technique. The scintillator detectors were shielded appropriately by lead and plastic blocks for the purpose of reducing noise level mainly originated by gamma rays and thermal neutrons hitting the scintillators. Two 5cm thick, one-meter long plastic collimators, with 10 cm $\times$ 10 cm internal aperture, were used along the line of sight of the scintillators in order to reduce stray neutrons hitting the detector. 

The three plastic scintillators discussed in this paper are EJ232Q\cite{EJ232} (0.5\% quenching, 150 mm diameter, 25 mm thick), BC422Q\cite{BC422} (1\% quenching, 180 mm diameter, $10$ mm thick) and EJ410\cite{EJ232} (90 mm diameter, 16 mm thick). Calibration of EJ232Q and EJ410 scintillators was obtained in the same set of shots while using the two detectors inside the detector housing as shown in the Fig.~\ref{fig:epsart1}. However, the BC422Q scintillator was calibrated with the BDS in a separate set of shots where the BC422Q detector was kept alone inside the detector housing. Where EJ232Q and BC422Q have fast rise and decay times  (sub ns and a few ns respectively), EJ410 has significantly slower decay of 200 ns. Each scintillator was coupled with a fast PMT tubes for converting the light output from the scintillator into electrical signal as well as to amplify the signal for detection. The gain of PMT was controlled by varying the biasing voltage from shot to shot in order to avoid signal saturation. The detectors were attached to fast (6GHz) oscilloscopes for recording the output from the PMTs.

\section{\label{sec:working} Basics of Scintillation process}

Since neutrons do not interact via the Coulomb force, they don't produce excitations in the scintillation material directly. Therefore neutron detection by scintillators relies on secondary ionising particle generation in the scintillator material via elastic, inelastic or nuclear reactions. Fast neutrons (generally $\gtrsim$ 0.5 MeV) can efficiently produce recoil protons via elastic scattering in hydrogen-rich materials, whereas slow neutrons rely on nuclear reactions such as the (n,$\gamma$) or (n,$\alpha$) reactions to produce ionisation. Materials such as LiI(Eu) or glass silicates doped with $^{6}$Li and $^{10}$B are therefore particularly well-suited for the detection of slow (thermal) neutrons.

The response of plastic scintillators to fast neutrons depends on several factors. The dominant mechanism for production of secondary ionising radiation for incident neutron energy up to a few 10s of MeV is the proton recoil based on elastic n-p scattering. As neutrons of a given energy pass through the scintillator, they produces a spectrum of recoiled protons by direct elastic scattering with hydrogen atoms present inside the material. Energy of the recoiled protons varies from zero to a maximum value, which is equal to the incident neutron energy. Each recoiled proton then produces scintillation light, $L_p(E)$, while travelling in the scintillator over a finite length depending on its kinetic energy. A part of the energy deposited by the protons is used for the scintillation process and the $L_p(E)$ can be calculated with semi-empirical formulae described by Chou\cite{o1996response} and Wright\cite{mouatassim1995light}. Where Chou's formula, given as 
\begin{equation}
\frac{dL_p(E)}{dx}=S \frac{dE}{dx}\left[1+KB\frac{dE}{dx}+C{\left(\frac{dE}{dx}\right)}^2\right]^{-1}
\label{Chouformula}
\end{equation}

which is a generalisation of the Birk's formula~\cite{birks1964theory}, Wright formula, 

\begin{equation}
L_p(E)=A \int^R _0 ln\left[1+\alpha\left(\frac{dE}{dx}\right)\right] dx,
\label{Wrightformula}
\end{equation}

on the other hand uses a simpler model while attributing for quenching effects in scintillators. 
In the above formulae, $S$, $KB$, $C$, $A$ and $\alpha$ are the fitting parameters to the experimental data, $R$ is range of the proton inside the scintillator and $dE/dx$ is the proton energy loss in the scintillator material per unit length. 

%%%%%%%%%%%%%%%%%%%%%%%%%%%%%%%%%%%%%%FIGURE%%%%%%%%%%%%%%%%%%%%%%%%%%%%
\begin{figure*}
\includegraphics[width=0.5\textwidth, angle =-90]{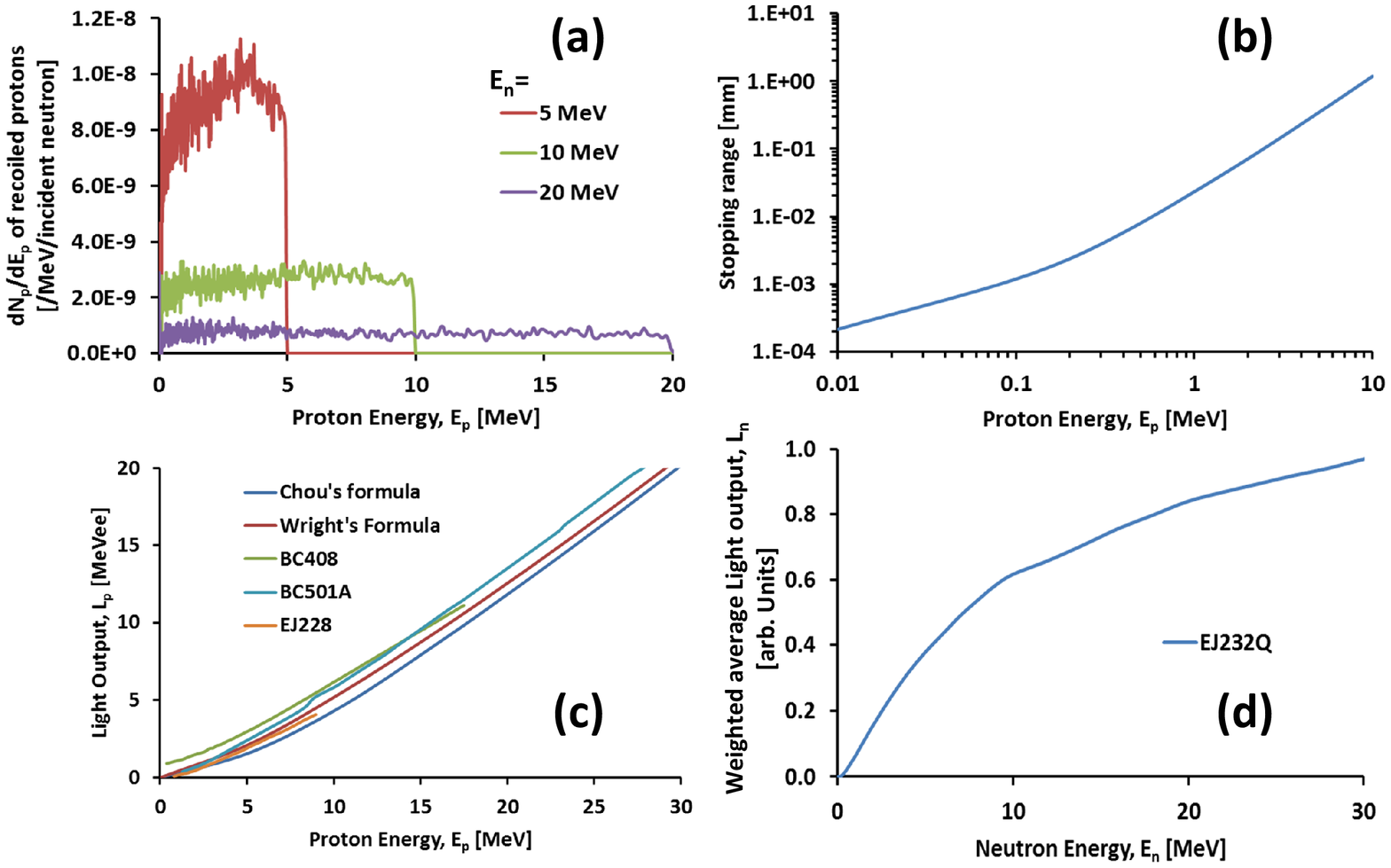}
\caption{\textbf{a)} Spectra of recoiled protons generated inside a $10\um\times 10\um\times 10\um$ cube of EJ232Q scintillator for different incident neutron energies. \textbf{b)} shows stopping range of protons inside EJ232Q scintillation material obtained by SRIM simulation. \textbf{c)} shows scintillation light output for different proton energies calculated for EJ232Q scintillator by using Chou's and Wright's formulas, as well as for a few different scintillators obtained from literature\cite{pozzi2004analysis}.
\textbf{(d)} shows the relative, weighted average light output from the EJ232Q obtained by using Eq.\ref{l_n}, where the recoiled proton spectrum is shown in (a) and $L_p(E)$ is shown in (c).}
\label{Figures_2}
\end{figure*}
%%%%%%%%%%%%%%%%%%%%%%%%%%%%%%%%%%%%%%%%%%%%%%%%%%%%%%%%%%%%%%%%%%%%%%%%%%%%%

\section{\label{sec2:leveliv} Relative Response of EJ232Q to fast neutrons}

Although Chou's and Wright's semi-empirical formulae provide fairly accurate estimation of the light output, the statistical nature of particle scattering and energy loss mechanisms demands a thorough monte-carlo approach in order to characterise the response of plastic scintillators to fast neutrons.  
Most of the commercially available plastic scintillators are experimentally calibrated for different types of ionising radiation, such as electrons, protons, alpha particles etc., as provided by the manufacturers. However the calibration of plastic scintillators for fast neutrons is often unknown, as is the case for EJ232Q and BC422Q used in our experiment. Therefore we used a  simple model based on a series of systematic monte-carlo simulations, as described below, in order to obtain a relative response in terms of the light output by the whole spectrum of recoiled protons.

At first the energy spectrum of recoiled protons produced by a given energy of neutrons in EJ232Q plastic (which is the same material used in BC422Q) was obtained by employing FLUKA\cite{ferrari2005fluka} simulations. The stoichiometric information about the EJ232Q plastic was taken from the manufacturer documentation. A series of FLUKA simulations was done by irradiating a small cube of the scintillator material ($10\um\times 10\um\times 10\um$) with different neutron energies. For each neutron energy the recoiled protons generated inside the cube were tallied. As an example, Fig.~\ref{Figures_2}\textcolor{red}{a} shows the recoiled proton spectra produced by neutrons of three different energies. 

The next step of calculation involves finding the stopping range of protons in the EJ232Q material, which was simulated using SRIM\cite{ziegler2010srim} monte-carlo simulation. The stopping range was calculated by obtaining $dE/dx$ as a function of incident energy of protons while passing through a very thin ($1~\um$) layer of EJ232Q, as shown in Fig.~\ref{Figures_2}\textcolor{red}{b}.  Using the stopping range of protons, $L_p(E)$ for the EJ232Q scintillator material was calculated for different proton energies by using Chou's and Wright's semi-empirical formulae, as shown in Fig.~\ref{Figures_2}\textcolor{red}{c}. As one can see, the $L_p(E)$ from both formulae closely agree with each other and with the data provided by the manufacturer. It also closely matches with the data found in literature for similar plastic scintillators\cite{pozzi2004analysis}. 

Finally, $L_n(E_n)$, the weighted average scintillation light output for a given neutron energy $E_n$ is obtained by integrating the light output from all the recoiled protons generated by the fast neutrons inside the scintillator. Mathematically one can write 
\begin{equation}
\label{l_n}
L_{n}(E_{n})=\int_0 ^{E_{max}=E_n} ~~\left.\frac{dN_{p}}{dE}\right|_{E_n}L_{p}(E)~dE
\end{equation}
where ${dN_p}/{dE}$ is the recoil proton spectrum shown in Fig.~\ref{Figures_2}\textcolor{red}{a} and $L_p(E)$ is the light output by different energy protons in the scintillator, shown in Fig.~\ref{Figures_2}\textcolor{red}{c}. By following the steps mentioned above, the weighted average light output from the scintillator was obtained for different incident neutron energies as shown in Fig.~\ref{Figures_2}\textcolor{red}{d}.

Since the BC422Q is the same scintillator material as EJ232Q, produced by a different company, we used the same detector response for EJ232Q and BC422Q in the data analysis described below. However, for the EJ410 scintillator we used the relative response of the scintillator to different neutron energy provided by the manufacturer~\cite{EJ232}.

%%%%%%%%%%%%%%%%%%%%%%%%%%%%%%%%%%%%%%%FIGURE%%%%%%%%%%%%%%%%%%%%%%%%%%%%%%
\begin{figure}[b]
\includegraphics[width=0.31\textwidth,angle=-90]{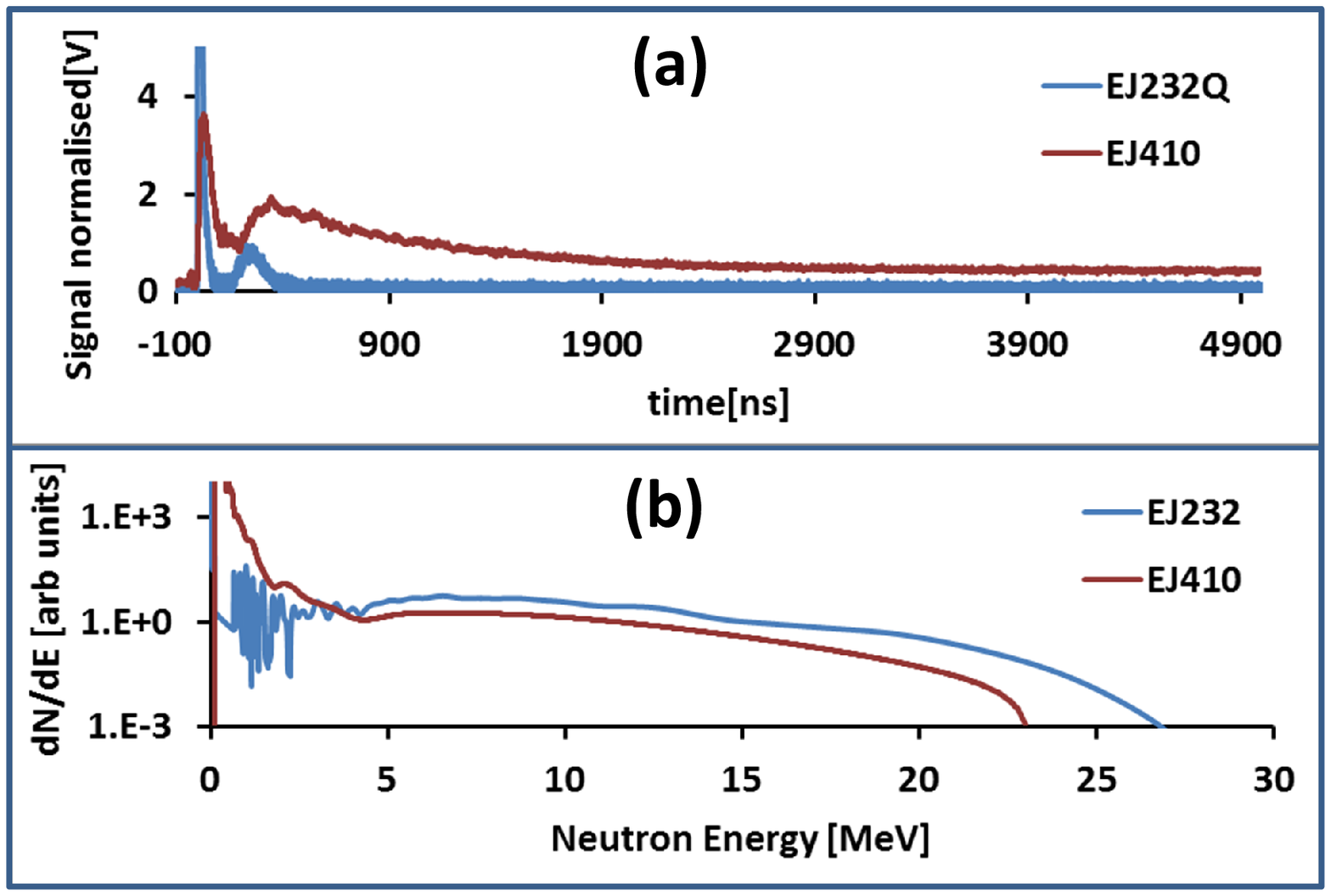}
\caption{\label{figure_3} \textbf{(a)} Typical raw data obtained by the two different scintillators, EJ232Q and EJ410, in the same shot. The sharper peak in the signal near to zero time corresponds to prompt gamma irradiation, while the following peak corresponds to neutron signal. \textbf{(b)} shows deconvolved neutron spectra from the raw signal shown in (a), after accounting for PMT gain and scintillator detection response.}
\end{figure}
%%%%%%%%%%%%%%%%%%%%%%%%%%%%%%%%%%%%%%%%%%%%%%%%%%%%%%%%%%%%%%%%%%%%%%%

\begin{figure}[b]
\includegraphics[width=0.31\textwidth, angle = -90]{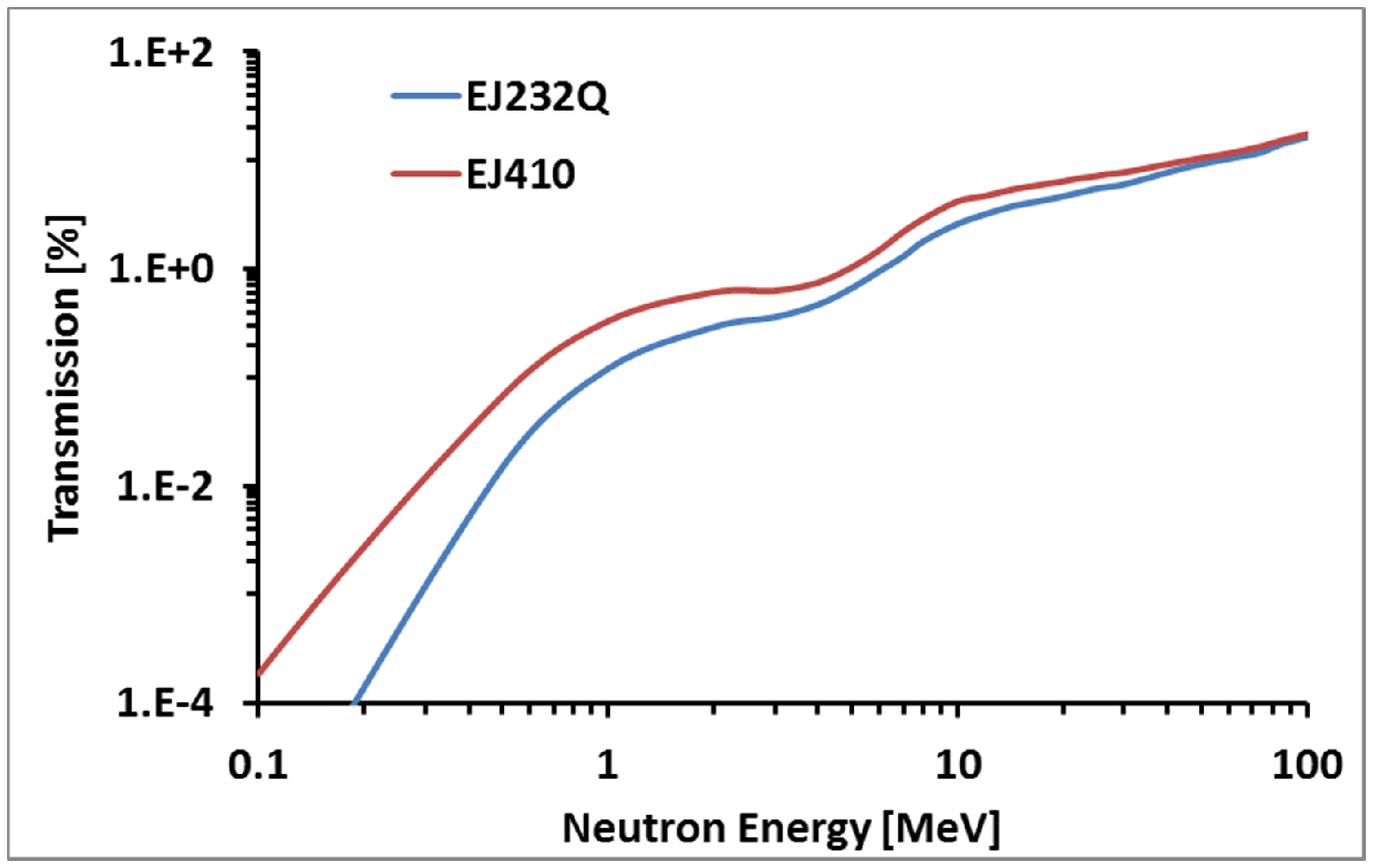}
\caption{\label{figure_4} Transmission in \% of different energy neutrons from source to the detector obtained by FLUKA simulations for the setup shown in Fig.~\ref{fig:epsart1}.}
\end{figure}

\section{\label{sec2:levelv} Calibration of the scintillator detectors}

\begin{figure*}
\includegraphics[width=0.31\textwidth, angle = -90]{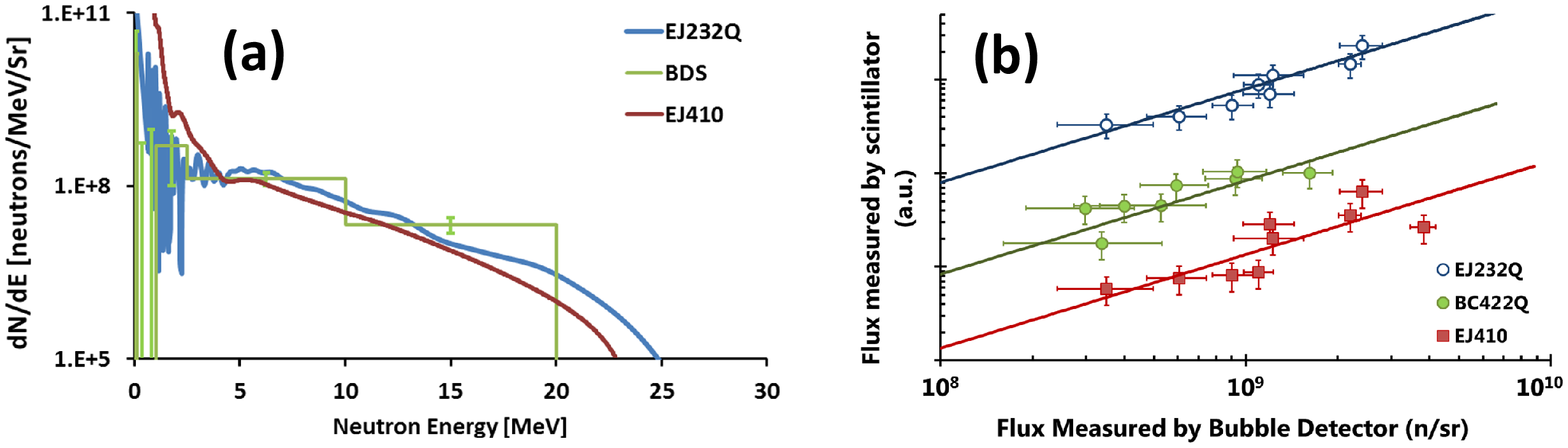}
\caption{\label{Figure_5} \textbf{(a)} Calibrated neutron spectra obtained from EJ232Q and EJ410 detectors compared with that obtained by BDS in the same shot. \textbf{(b)} Comparison between neutron flux obtained from different scintillator detectors and the BDS for neutron energy in the range 2.5-20 MeV. The graph plots the comparison for three different scintillator detector, EJ232Q, BC422Q and EJ410, obtained over a number of shots. The solid lines show the linear fit to the data points which passes through origin.}
\end{figure*}

The TOF arrangement allows us to distinguish the signal produced by the neutrons from the gamma-rays, produced during the laser interaction, by their difference in arrival time at the detector. The prompt gamma flash in the raw data at early time, as can be seen in the Fig.~\ref{figure_3}\textcolor{red}{a}, serves as a precise temporal fiducial to identify the time of neutron generation at the source. Using this temporal reference, the time axis of the raw data was converted into neutron energy by taking into account the distance between the source and the detector. Due to the large distance ($\sim$11 meter) between the source and the detector employed in the experiment, the signals due to gamma radiations and neutrons are well separated in time, as can be seen in Fig.~\ref{figure_3}\textcolor{red}{a}, which allowed us to discard unambiguously the gamma signal from the raw data before processing further.

The scintillator detectors used in the experiment were coupled with fast photo-multiplier tubes in order to convert the optical photons from the scintillator into electrical signal with a gain commensurate to the applied bias voltage. The gain characteristics of the PMTs are calibrated precisely by the manufacturers, which were used in the data analysis, following the background subtraction of the raw data and deconvolution of the signal by scintillation temporal profile. Assuming the light collection by detector geometry, transmission through light guide, quantum efficiency of the PMT etc. are constants for a given detector, each point in the processed data now corresponds to the light emitted by the scintillator due to the neutrons within the respective energy bin.  
Since the energy bins are typically very small (for example, $\sim$1keV at 1 MeV and $\sim$10 keV at 10 MeV), the relative number of neutrons in a given energy bin, $dN(E_n)$, is now obtained by dividing each point of the data by the corresponding $L_n(E_n)$ (given in Fig.~\ref{Figures_2}\textcolor{red}{(d)}). Finally, the neutron spectrum (dN/dE) was obtained by dividing each data point by the width of the respective energy bin. Fig.~\ref{figure_3}\textcolor{red}{b} shows neutron spectrum obtained from the data collected by the EJ232Q and EJ410 detectors. However, the spectrum does not yield the absolute number of neutrons  hitting the detector because of the various unknown factors, such as light collection by the detector housing, attenuation of scintillation photon by the light guide, quantum efficiency of the PMT photocathode etc. Therefore, it is important to cross-calibrate the detectors against a standard dosimeter. In our case, absolute flux calibration was carried out against BDS fielded in the same shot along the same direction of observation. 

Bubble detectors (BD) are passive detection devices that provide neutron response proportional to the number of bubbles formed in the detector\cite{birnboim1984bubble}. The response of the bubble detectors depends on the kinetic energy of the incident neutrons and the composition of the BD medium\cite{lewis2012review}. Therefore the BDS consists of six sets of bubble detectors calibrated for different energy ranges of neutrons. After exposure to the broadband neutron source, the number of bubbles in each of the bubble detector was manually counted and the neutron spectrum is obtained by deconvolving the number of bubbles in each set of detectors against their calibration curves provided by the manufacturer.

In order to cross-calibrate the scintillator detectors against the data obtained in the BDS, it is important to assess the transmission and scattering of neutrons by the various objects (including the BDS) they encounter along the scintillators' line of sight. Since the scintillators are placed at a significantly large distance ($\sim$ 11 meter) from the BDS, any small angle scattering of neutrons by the intermediate objects would prohibit the scattered neutrons from reaching the scintillator detector. On this basis the neutron propagation through the system was simulated using FLUKA in order to estimate transmission to the detector as a function of neutron energy, as shown in the figure~\ref{figure_4}.

In the simulation, the neutron source was defined as a pencil beam with no divergence. The simulation was set by considering all different objects on the line of sight to the scintillator detector, including the chamber wall and the detector housing. A series of simulations were carried out for different neutron energies at the source and the neutron spectra at the corresponding detector was tallied. Due to the heavy shielding (typically 5 cm of lead and 5 cm of plastic) used around the detectors (to suppress the noise level in the detector by gamma rays and thermal neutrons) the scattered neutrons in the vicinity of the detectors can in principle contribute towards the signal for that energy in a TOF mode. In order to account for the scattered neutrons, two sets of simulations were done for each energy, one with and one without the lead and plastic shielding around the detector. The neutron spectra at the detector obtained from these two simulations were subtracted in order to estimate the spectra of scattered neutrons reaching the detector due to the detector housing. The transmission factor is then calculated by taking into account the scattered neutron spectra and the number of neutrons of the input energy, convolved with the neutron detection response of the detector. 

The neutron transmission calculated above was used in the TOF data analysis in order to obtain the neutron spectrum at the BDS plane from the neutron spectrum at the scintillator plane, as shown in Fig.~\ref{figure_3}\textcolor{red}{b}. In this way, a direct comparison between neutron flux obtained by the scintillator and BDS can be acquired for an absolute calibration. A typical data set is shown in fig.~\ref{Figure_5}\textcolor{red}{a} where the neutron spectra recorded from EJ232Q and EJ410 scintillators are shown in comparison with the spectrum obtained by BDS, after applying a constant factor in order to match the neutron flux with that measured by the BDS. One can see a good agreement between the spectra in terms of spectral shape, which remains fairly consistent. 

The comparison between the neutron flux obtained from different scintillators with the one measured absolutely by the BDS over a number of shots is shown in figure~\ref{Figure_5}\textcolor{red}{b}. In this case we have chosen the high energy range (2.5-20 MeV) for flux comparison in order to avoid discrepancies due to the down-scattered (lower energy) neutrons detected by the BDS, which is a time integrated spectrometer. As can be seen from the fig.~\ref{Figure_5}\textcolor{red}{b}, the data points follow a linear trend over a range of neutron fluxes, the slope of which can be used as the final calibration factor for the scintillators after applying other corrections, such as the scintillator detection response, the PMT gain and the transmission function for the setup. 

\section{\label{sec4:levelc}Conclusion}% 

Employing a sub-ns neutron source driven by high power laser and absolutely calibrated, gamma insensitive bubble detector spectrometers, we have carried out the systematic calibration of three types of plastic scintillators namely EJ232Q, BC422Q and EJ410. Cross-calibration of the scintillators were done by implementing the neutron detection response (modeled semi-empirically for EJ232Q, BC422Q and using the one obtained from the manufacture for EJ410), PMT gain factors obtained from the manufacturers and transmission of neutron through our experimental setup as simulated by FLUKA. Over a number of shots, spanning over a wide range of neutron fluxes between $10^8 - 10^{10}$ n/Sr for 2.5-20 MeV neutrons, the comparison between the results obtained from the TOF data analysis and the BDS data follows a linear fit, which provides the final calibration factor for the scintillators.  

\section*{Acknowledgements}\label{sec:Acknowledgement}
The authors acknowledge funding from EPSRC [EP/J002550/1-Career Acceleration Fellowship held by S. K., EP/L002221/1, EP/E035728/1, EP/K022415/1, EP/J500094/1 and EP/I029206/1], Laserlab Europe (EC-GA 284464), projects ELI (Grant No. CZ.1.05/1.1.00/483/02.0061) and OPVK 3 (Grant No. CZ.1.07/2.3.00/20.0279). Authors also acknowledge the support of ESG, mechanical and target fabrication staffs of the Central Laser Facility, STFC, UK.

\nocite{*}
%\section{\label{sec:level1}References \lowercase{} }
\providecommand{\noopsort}[1]{}\providecommand{\singleletter}[1]{#1}%

\end{document}